\documentclass[12pt]{article}
\usepackage{amsmath,amsthm,amssymb}
\begin{document}
\newtheorem{proposition}{Proposition}[section]
\newtheorem{corol}[proposition]{Corollary}
\newtheorem{lemma}[proposition]{Lemma}
\newtheorem{claim}[proposition]{Claim}
\theoremstyle{definition}
\newtheorem{definition}[proposition]{Definition}
\theoremstyle{remark}
\newtheorem{example}[proposition]{Example}
\newtheorem{remark}[proposition]{Remark}
\renewcommand{\emptyset}{\varnothing}
\renewcommand{\leq}{\leqslant}
\renewcommand{\geq}{\geqslant}
\title{Solving equations in the relational algebra}
\author{Joachim Biskup \and Jan Paredaens \and Thomas Schwentick \and 
Jan Van den Bussche\thanks{Contact author. Address: Limburg University (LUC),
B-3590 Diepenbeek, Belgium. Tel: +32-11-268226. Fax: +32-11-268299. Email:
jan.vandenbussche@luc.ac.be.}}
\date{}
\maketitle
\begin{abstract}
Enumerating all solutions of a relational algebra equation
is a natural and powerful operation which, when added as a
query language primitive to the nested relational algebra,
yields a query language for nested relational databases,
equivalent to the well-known powerset algebra.  We study
\emph{sparse} equations, which are equations with at most
polynomially many solutions.  We look at their complexity,
and compare their expressive power with that of similar
notions in the powerset algebra.
\end{abstract}

\section{Introduction}
Suppose we are allowed to see only a view on a database $B$, computed by
a relational algebra expression $e$. If we still want to find out what
$B$ is, we might try to ``invert'' $e$ (assuming we know this
expression), which will only work when we also know the finite domain
$D$ of $B$.  Specifically, we can enumerate all databases $X$ over $D$,
and test for each $X$ whether it satisfies the equation $e(X) = e(B)$.
One of these solutions will be $B$ of course, so if the set of all
solutions is not too big, it might provide us with useful information
to start our detective work.

The above simple scenario from database security led us to wonder what can be
said in general about the solution of equations in the relational algebra.
Generally, if $e_1$ and $e_2$ are two algebra expressions over some
database schema augmented with some relation variables $X_1$, \dots,
$X_p$, we can consider the equation $e_1 = e_2$.  A
\emph{solution} of this equation, given a database $B$ with finite
domain $D$, is a tuple $(X_1,\dots,X_p)$ of relations over $D$ such that
$e_1$ and $e_2$ evaluate to the same relation on the augmented database
$(B,X_1,\dots,X_p)$.

Asking whether there exists a solution of a relational algebra
equation on a database is almost exactly the same thing as asking whether an
existential second-order logic sentence is true on that database.
Hence, by Fagin's theorem \cite{fagin_theorem,ef_fmt},
the problems that can be formulated as
finding a solution of some relational algebra equation are nothing but
the problems in NP.

However, in the present paper, we start from the observation that the
set of all solutions of an equation, being a set of tuples of relations,
is a \emph{nested} relation.  One can therefore consider the enumeration
of all solutions of an equation as a query language primitive, which can
be added to the nested relational algebra.  We introduce and study this
extension of the nested relational algebra, which we call the
\emph{equation algebra}. The equation algebra is extremely powerful: it
is equivalent to the well-known powerset algebra for nested relations.
Our particular interest, however, is in what can be expressed in the
equation algebra by using only equations that have a solution set of
polynomial size on each database.  We call such equations \emph{sparse.}

Our interest in sparse equations does not stem from time efficiency
considerations.  Indeed, it is not obvious how knowing that an equation
is sparse would help in actually finding even one solution more quickly.
It is neither obvious, however, that it would \emph{not} help.  For
example, consider the problem of checking on a given database whether
some fixed sparse relational algebra equation has a solution.
Using an extension of Fagin's theorem to nested relational databases, we show
that
this problem can be NP-hard only if every problem in NP can
already be decided by a polynomial-time non-deterministic Turing machine
that has only polynomially many accepting computations on each input.
The latter is one of the many unresolved questions in computational
complexity theory \cite{allender_sparsep}.

Nevertheless, sparse equations are still interesting from a space
efficiency standpoint.  Indeed, for the natural evaluation strategy for
equation algebra expressions to run in polynomial space, it is necessary
that all equations
occurring in the expression are sparse.  Interest in fragments of
powerful query languages for which the natural evaluation strategy is
polynomial-space is not new to database theory research. For example,
Abiteboul and Vianu \cite{av_generic} showed that the parity query is
not expressible in the polynomial-space fragment of various
computationally complete query languages.

Closer to our topic is the work of Suciu and Paredaens
\cite{sp_powerset}, who showed that queries such as transitive closure
and parity are not expressible in the polynomial-space fragment of the
powerset algebra for nested relations. This fragment
consists of all powerset algebra expressions where all intermediate
results are of polynomial size, on each database. Note that this fragment does
make sense as there are expressions that always produce a result of
logarithmic size;  applying the, exponential, powerset operator to such
expressions produces a result of polynomial size. 

We also mention
Grumbach and Vianu \cite{gv_tractable}, who also studied a sparsity
notion in connection with queries over nested relational databases,
although they considered sparsity as a property of databases rather than
of query language expressions.

Suciu and Paredaens conjectured in general that the polynomial-space
fragment of the powerset algebra has no more power than the nested
relational algebra without powerset.  (This conjecture has been confirmed for
monadic database schemas \cite{vdb_ff}.) At first sight, the operator
that we add to the nested relational algebra, to enumerate all solutions
of an equation, does not seem to be that different from the powerset
operator. After all, both operators
perform some kind of potentially exponential enumeration.

Yet, as we will point out, the analogue of the Suciu-Paredaens conjecture
does not
hold for the sparse fragment of the equation algebra.  Specifically,
using sparse equations only, we can express transitive closure; in fact,
we can express any fixpoint query. This complements a result by
Abiteboul and Hillebrand \cite{ah_space}, who showed that transitive
closure becomes expressible in the powerset algebra in polynomial space,
provided we use a more clever ``pipelined'' evaluation strategy.
Actually, every fixpoint query is already expressible using equations
that are not just sparse, but even unambiguous: they have a
unique solution on each database. Unambiguous equations in the
relational algebra are known as \emph{implicit definitions} in
first-order logic, and were studied in the context of finite model
theory by Kolaitis \cite{kolaitis_implicit}.  Kolaitis already showed
that every fixpoint query can be implicitly defined.  We offer a
straightforward, direct proof.

Another example of the differences between the sparse fragment of the
equation algebra and that of the powerset algebra
is given by the well-known nesting operator of
the nested relational algebra.  This operator becomes redundant once we
extend the algebra with the solution operator or with the powerset
operator.  However, the original nesting operator
never blows up exponentially. We show that, without using the nesting operator
itself, nesting is expressible 
in the equation algebra using sparse equations only,
but that the same is not possible with a
polynomial-space powerset algebra expression.

However, there are also similarities between the two fragments.
Specifically, we prove an analogue to the Suciu-Paredaens result, to the
effect that the parity query is not expressible in the sparse
fragment of the equation algebra either.
This is our main technical contribution;
the proof
is in the style of  an elegant argument of Liebeck \cite{liebeck},
invoking Bochert's theorem on the order of primitive permutation groups.
Coming back to
connection with implicit definitions in first-order logic,
our result generalizes the known and easy fact
that the parity query cannot be defined
implicitly \cite{kolaitis_implicit},
in two directions: from unambiguous to
sparse; and from a single equation
to an arbitrarily complex expression involving
several, possibly nested, equations.

This paper is organized as follows.  Section~\ref{secprelim} recalls the
nested relational data model.  Section~\ref{secequations} introduces
relational algebra equations.  Section~\ref{secalgebra} introduces the
equation algebra.  Section~\ref{secsparse} introduces sparse equations, as
well as the natural evaluation strategy for equation algebra expressions.
Section~\ref{sectime} studies the time complexity of sparse equations.
Finally, Section~\ref{secresults} presents the comparison with the
polynomial-space powerset algebra.

\section{Preliminaries} \label{secprelim}

We quickly recall the nested relational data model and algebra
\cite{thomfisch,ahv_book}.

\emph{Relation types} are defined as follows.
The symbol $0$ is a type; and, if
$\tau_1$, \dots, $\tau_k$ are types, then so is $(\tau_1,\dots,\tau_k)$.
For a type $\tau$, and some set $D$ of atomic values, the \emph{relations of
type $\tau$ on $D$} are inductively defined as follows.
A relation of type $0$ on $D$ is just an
element of $D$ (this serves merely as the base case for the 
induction).  A relation of type $(\tau_1,\dots,\tau_k)$ on $D$ is a set of
$k$-tuples $(x_1,\dots,x_k)$ such that $x_i$ is a relation of type $\tau_i$
on $D$, for $i=1,\dots,k$.

A \emph{database schema} is a finite set $\cal S$ of relation names, where each
relation name has an associated type different from $0$.
A \emph{database $B$} over $\cal S$ consists of a non-empty
finite domain $D$ of atomic values, together with, for each
relation name $R$ in $\cal S$, a relation $R^B$ of type $\tau$ on $D$, where
$\tau$ is the type of $R$.

The operators of the nested relational algebra are those of the standard
relational algebra (union $\cup$ and difference $-$
of relations of the same type;
cartesian product $\times$; projection $\pi$; selection $\sigma$
for equality, which can now be set equality of nested
relations), plus the operators
\emph{nesting} $\nu$ and \emph{unnesting} $\mu$, defined as follows.

Let $R$ be a relation of type $(\tau_1,\dots,\tau_k)$, and let 
$i_1,\dots,i_p \in \{1,\dots,k\}$.  Then the nesting
$\nu_{i_1,\dots,i_p}(R)$ equals the relation
\begin{multline*}
\Bigl\{\Bigl(x_1,\dots,x_k,
\bigl\{(y_{i_1},\dots,y_{i_p}) \mid
(y_1,\dots,y_k) \in R \\
\text{and } x_j = y_j \text{ for each }
j \in \{1,\dots,k\} - \{i_1,\dots,i_p\} \bigr \} \Bigr ) \\
{} \mathrel{\Bigl |} (x_1,\dots,x_k) \in R \Bigr \}
\end{multline*}
of type $(\tau_1,\dots,\tau_k,(\tau_{i_1},\dots,\tau_{i_p}))$.

Let $R$ be as in the previous paragraph,
and let $i \in \{1,\dots,k\}$ such that $\tau_i \neq 0$;
so $\tau_i$ is of the form $(\omega_1,\dots,\omega_\ell)$.
Then the unnesting $\mu_i(R)$ equals the relation
\[
\{(x_1,\dots,x_k,y_1,\dots,y_\ell) \mid {} 
(x_1,\dots,x_k) \in R \text{ and }
(y_1,\dots,y_\ell) \in x_i\}
\]
of type
$(\tau_1,\dots,\tau_k,\omega_1,\dots,\omega_\ell)$.

The expressions of the \emph{nested relational algebra} over a
schema $\cal S$
are now built up using the above operators from the relation names in $\cal S$
and the symbol $D$, which stands for the finite domain of the input database.
The relation to which an expression $e$ evaluates on a database $B$ is denoted
by $e(B)$.

One can extend the nested relational algebra to the \emph{powerset algebra} by
adding the powerset operator, defined as follows.  Let $R$ be a relation of
type $(\tau_1,\dots,\tau_k)$.  Then the powerset $\Pi(R)$ equals the relation
$\{S \mid S \subseteq R\}$ of type $((\tau_1,\dots,\tau_k))$.

\section{Equations} \label{secequations}

Let $\cal S$ and $\cal X$ be disjoint database schemas; 
$\cal S$ is the actual database schema, while 
$\cal X$ is thought of as a set of additional relation variables.
Let $e_1$ and $e_2$ be two
expressions over the expanded schema ${\cal S} \cup {\cal X}$.

\begin{definition} \label{defsolution}
Given a database $B$ over $\cal S$, a \emph{solution to the equation
$e_1=e_2$} is a database $A$ over $\cal X$ with the same finite domain as $B$,
such that $e_1(B,A) = e_2(B,A)$.
\end{definition}

Here, $(B,A)$ denotes the expansion of $B$ with $A$, i.e., the database over
${\cal S} \cup {\cal X}$ that has the same finite domain as $B$, that equals
$B$ on ${\cal S}$, and that equals $A$ on $\cal X$.

\begin{example} \label{solex}
For a very simple example,
let $R \in {\cal S}$ and let ${\cal X} = \{X\}$, where $X$ has the same type
as $R$.  Then $X \cup R = R$ is an equation.  Given a database $B$ over $\cal
S$, a database $A$ over $\{X\}$ is a solution if and only if $X^A \subseteq
R^B$.

For another example,
let $X$ be a relation variable of type $(0,0)$.
One can write a relational algebra expression $e$
such that on any database $A$ over $\{X\}$ with
finite domain $D$,
$e(A)$ is empty if and only if $X^A$ is one-to-one, the projections
$\pi_1(X^A)$ and $\pi_2(X^A)$ are disjoint, and their union equals $D$.
An example of an $e$ that works is
\begin{multline*}
\pi_1 \sigma_{2 \neq 4} \sigma_{1=3}(X \times X)
\cup \pi_2 \sigma_{2=4} \sigma_{1 \neq 3}(X \times X) \\
\cup \bigl ( \pi_1(X) - (\pi_1(X) - \pi_2(X)) \bigr ) \\
{} \cup \bigl (D - (\pi_1(X) \cup \pi_2(X)) \bigr ) \cup
\bigl ((\pi_1(X) \cup \pi_2(X)) - D \bigr )
\end{multline*}
Then the equation $e = \emptyset$ has a solution
on a database $B$ with finite domain $D$ if and only
if the cardinality of $D$ is even.  (Technically,
$e=\emptyset$ is not an equation because the symbol $\emptyset$ is not an
expression, but we can easily take $\emptyset$ here to stand for the
expression $D-D$ which always evaluates to the empty relation.)
\qed
\end{example}

\begin{remark} \label{disequation}
In the above example, we used an equation of the special form $e=\emptyset$.
Actually, this form is not so special at all, because any equation $e_1 = e_2$
can be brought in this form as $e_1 \mathbin{\Delta} e_2 = \emptyset$, where
$e_1 \mathbin{\Delta} e_2$ stands for $(e_1 - e_2) \cup (e_2 - e_1)$
(symmetric difference).

Alternatively, one might wonder about the use of \emph{dis}equations, of the
form $e \neq \emptyset$.  These are nothing but equations in disguise,
because they can also be written as $\pi_1(D \times e) = D$.
Conversely, any equation $e_1 = e_2$ can also be written as the disequation
$D - \pi_1(D \times (e_1 \mathbin{\Delta} e_2)) \neq \emptyset$.
\qed
\end{remark}

\section{The equation algebra} \label{secalgebra}

We are now ready to extend the nested relational algebra with a
solution operator for equations. 
We refer to the resulting algebra as the \emph{equation algebra.}

To allow for an elegant definition, we do not fix a schema $\cal S$ in
advance.  Rather, we assume a sufficiently large supply of
relation names of all possible types.  Any relation name can now occur
in an expression. Like in logic formulas, some will occur \emph{free}
and others will occur \emph{bound.}  Bound relation names are bound by
the solution of an equation, and serve as the
variables of the equation.  Within the equation, however, they are still
free.  We denote the set of relation names that occur free in an
equation algebra expression $e$ by ${\it free}(e)$.

For the constructs of the nested relational algebra, this is all
straightforward:  for a relation name $R$, we have
${\it free}(R) := \{R\}$; for expressions $e$ of the
form $(e_1 \cup e_2)$, $(e_1 - e_2)$, or $(e_1 \times e_2)$, 
we have ${\it free}(e) := {\it free}(e_1) \cup {\it free}(e_2)$; for
expressions $e$ of the form $\sigma(e')$, $\pi(e')$, $\nu(e')$, or $\mu(e')$,
we have ${\it free}(e) := {\it free}(e')$.  For the expression $D$, we have
${\it free}(D) := \emptyset$.

The definition of the new solution operator is now the following:
\begin{definition}
Let $e_1$ and $e_2$ be expressions,
and let $X_1,\dots,X_p$ be a sequence of distinct relation names.
Then $$ \{(X_1,\dots,X_p) \mid e_1 = e_2\} $$
is also an expression (called a solution expression).
We define its $\it free$ set as
$({\it free}(e_1) \cup {\it free}(e_2)) - \{X_1,\dots,X_p\}$.
We say that the $X_i$ \emph{become bound.}
\end{definition}

Note that this is a recursive
definition, in the sense that $e_1$ and $e_2$ can contain solution operators
in turn.  To avoid clutter, we disallow equation algebra expressions in
which a free relation name at the same time becomes bound in some
subexpression, as in $X \times \{(X) \mid X \cup R = R\}$.

An expression $e$ in the equation algebra can be evaluated on databases $B$
over any schema that contains ${\it free}(e)$.  We already know how this
evaluation is defined for the constructs of the
nested relational algebra.  So we only have to give

\begin{definition}
For a solution
expression, $e$, of the form $\{(X_1,\dots,X_p) \mid e_1 = e_2\}$,
and a database $B$, the evaluation $e(B)$ equals the relation
\begin{multline*}
\{(X_1^A, \dots, X_p^A) \mid
\text{$A$ is a database over $\{X_1,\dots,X_p\}$} \\
\text{that is a solution of $e_1 = e_2$, given $B$}\}.
\end{multline*}
This relation is of type $(\tau_1,\dots,\tau_p)$, where $\tau_i$
is the type of $X_i$ for $i=1,\dots,p$.
\end{definition}

\begin{example} \label{exexpressions}
Recall the simple example equation $X \cup R = R$ from Example~\ref{solex}.
We can turn this equation in the following equation algebra expression $e$:
$\{(X) \mid X \cup R = R\}$, or, more readibly,
$\{(X) \mid X \subseteq R\}$.  Note that ${\it free}(e) = \{R\}$.
On any database $B$ over $\{R\}$, the relation $e(B)$ equals $\Pi(R^B)$
(recall the powerset operator $\Pi$ from Section~\ref{secprelim}).
In other words, the equation algebra expression $e$ is equivalent to the
powerset algebra expression $\Pi(R)$.

The equation algebra allows equations to be used inside equations.
For example, if we want to compute the powerset of the powerset of $R$, we can
write: 
$$ \bigl \{ (Y) \mid Y \subseteq \{ (X) \mid X \subseteq R \} \bigr \}. $$

As a third example, let $R$ and $T$ be relation names of the same
binary type $(\tau,\tau)$ for some $\tau$.
One can write a relational algebra expression $e_{\rm tc}$
such that on any database $C$ over $\{R,T\}$, $e_{\rm tc}$ is empty if and
only if $R^C \subseteq T^C$ and $T^C$ is transitively closed.
One can also write a nested relational algebra expression $e_{\rm min}$ 
that selects, out of a set of binary relations, the minimal ones w.r.t.\ set
inclusion.
Explicit forms for $e_{\rm tc}$ and $e_{\rm min}$ have been given by
Gyssens and Van Gucht \cite{gvg_powerset}.
Then the following equation algebra expression computes the transitive closure
of relation $R$:
$$ \pi_{2,3} \mu_1 e_{\rm min} \bigl ( \{(T) \mid e_{\rm tc} = \emptyset\}
\bigr ). $$
Indeed, the subexpression $\{(T) \mid e_{\rm tc} = \emptyset\}$ returns the
collection of all transitively closed relations on the same domain as $R$ and
containing $R$; applying $e_{\rm min}$ to that collection results in the
singleton consisting of the minimal element,
i.e., the transitive closure of $R$
(by definition of transitive closure); applying unnesting $\mu_1$ produces the
actual tuples in the transitive closure, keeping the nested relation (cf.~our
definition of the effect of $\mu$ in Section~\ref{secprelim}); and
applying $\pi_{2,3}$ finally removes the nested relation. \qed
\end{example}

In the above example we saw that the powerset operator is expressible in the
equation algebra.  Conversely, the solution operator is easily expressed in
the powerset algebra.  Hence,

\begin{proposition} \label{proppowerset}
The equation algebra is equivalent to the powerset algebra.
\end{proposition}
\begin{proof} To see that
$\{(X_1,\dots,X_p) \mid e_1 = e_2\}$ can be expressed in the powerset algebra,
we begin by noting that for any relation type $\tau$ one can write a powerset
algebra expression $\Pi^\tau$
yielding the collection of all relations of type $\tau$ on
$D$.  For example, $\Pi^{(0,0)}$ is $\Pi(D \times D)$,
and $\Pi^{(0,(0))}$ is
$\Pi(D \times \Pi(D))$.  Hence,
if the type of $X_i$ is $\tau_i$ for $i=1,\dots,p$,
then $\Pi^{\tau_1} \times \cdots \times \Pi^{\tau_p}$ yields the collection of
all potential solutions.  Now it suffices to observe that one can write nested
relational algebra expressions that
apply $e_1$ or $e_2$ to each database in this collection separately.
Explicit forms of such expressions have been given by Gyssens and Van Gucht
\cite{gvg_powerset}.  After that, the actual solutions can be selected by
an equality selection.
\end{proof}

\section{Sparse equations} \label{secsparse}

So far, the equation algebra is merely another syntax for
the powerset algebra, or, if you want, higher-order logic.
However, when we consider a natural
evaluation strategy for equation algebra
expressions, we start to notice some differences.

By the natural strategy to evaluate a solution expression of the form
$\{(X_1,\dots,X_p) \mid e_1=e_2\}$, we mean the following.  Enumerate
all databases $A$ over $\{X_1,\dots,X_p\}$, on the finite domain
of the given input database,
\emph{one by one, reusing the same space.}  For each $A$ we test
whether it is a solution (by recursively evaluating $e_1$ and $e_2$),
and if so, we include it in the result.

For the constructs of the nested relational algebra,
%or even the powerset algebra,
the natural evaluation strategy is clear: if we have
to evaluate an expression of the form $e_1 \mathbin{\Theta} e_2$, with
$\Theta \in \{\cup,-,\times\}$, we create two intermediate results by
recursively evaluating $e_1$ and $e_2$, and then apply $\Theta$ to these
two intermediate results.  Similarly, if we have to evaluate an
expression of the form $\theta(e)$, with $\theta \in
\{\pi,\sigma,\nu,\mu\}$ (and parameters added in subscript),
we create an intermediate result by
recursively evaluating $e$, and then apply $\theta$ to this intermediate
result.

In view of this natural evaluation strategy, we now propose

\begin{definition} \label{defsparse}
An equation is called \emph{sparse} if all its relation variables are of
\emph{flat} type, i.e., of type of the form $(0,\ldots,0)$,
and the number of solutions on any given database
is at most polynomial in the size of that database.
\end{definition}

\begin{example}
The two equations from Example~\ref{solex} are not sparse.
Probably the simplest example of a non-trivial
sparse equation is the following.
Let $X$ be a relation name of type $(0)$.
One can write a relational algebra expression $e$ over $\{X\}$ such that on 
any database $A$ over $\{X\}$, $e(A)$ is empty if and only if $X^A$ is a
singleton.  Then the equation $e=\emptyset$,
where $X$ is taken as the relation variable
to be solved for, is sparse.  Indeed, given any database $B$ with finite
domain $D$, the solutions are precisely all singleton subsets of $D$.
There clearly are only a linear (and thus at most polynomial)
number of possible solutions.
\qed
\end{example}

\begin{remark} \label{biskup}
A natural alternative definition of sparsity would be the one where, in
Definition~\ref{defsparse}, we would look only at databases over the schema
consisting of the relation names that actually occur free in the equation.
One easily sees, however,
that this alternative definition yields the same notion of sparsity.
\qed
\end{remark}

Sparse equations are connected to the natural evaluation strategy in the
following way:

\begin{proposition} \label{propsparse}
The natural strategy to evaluate an equation algebra expression $e$ runs in
polynomial space, if and only if all equations occurring in $e$ are
sparse.
\end{proposition}

Here, we count not only the space occupied by the
intermediate results stored during evaluation, but also the
size of the final result.

\begin{proof}
The if-direction is clear.  For the only-if direction,
we work by induction on the nesting depth of equations.
The base case---expressions that do not contain any equations at all---is
trivial.

For the inductive step, consider a top-level equation
$\{(X_1,\dots,X_p) \mid e_1=e_2\}$ occurring in $e$.
The natural strategy to evaluate this
equation runs in polynomial space, so in particular, for each expansion
of each database $B$ over ${\it free}(e)$ with a candidate solution $A$ over
$\{X_1,\dots,X_p\}$, the natural evaluation of $e_1$ and $e_2$ on $(B,A)$
runs in polynomial space.  In this way we consider every possible database
$C$ over ${\it free}(e) \cup \{X_1,\dots,X_p\}$, because the restriction of
$C$ to ${\it free}(e)$ is a possible $B$, and $C$ itself then is a possible
expansion of $B$.  Hence,
the natural evaluation strategies of $e_1$ and $e_2$ in general
run in polynomial space.

Formally, we must note here
that $e_1$ and $e_2$ might not
actually mention certain relation names in ${\it free}(e)$ or
$\{X_1,\ldots,X_p\}$, and that there is still the formal possibility
that their natural evaluation might not run in polynomial space on databases
over schemas not containing these names.  However, using Remark~\ref{biskup},
it can be seen that this is impossible.

By induction, we can therefore conclude that
all equations occurring nested
inside a top-level equation are sparse.

The top-level equation itself must also be sparse.  In proof, if one of the
$X_i$ would be of non-flat type, even one candidate solution can already be of
exponential size.  Indeed, even in the simplest case where $X_i$ would be of
type $((0))$, on a domain with $n$ elements, a possible
candidate value for $X_i$ 
is the collection of all subsets of that domain, which is of size
$2^n$.  So, every $X_i$ is of flat type.
Furthermore, since we store the solution set as an
intermediate result, it must be of at most
polynomial size on all databases over ${\it
free}(e)$.  Since the individual solutions are flat databases and thus of
polynomial size, the cardinality of the solution set must therefore be 
at most polynomial.
\end{proof}

\section{Time complexity of equation nonemptiness} \label{sectime}

The \emph{time} complexity of solving sparse equations is closely linked to 
an open question from computational complexity theory.  Unlike the previous
section, in this section we are
not talking about the natural evaluation strategy, whose time complexity is
clearly at least exponential as soon as there are equations to be solved.

Instead, we will be looking at the time complexity of the
\emph{nonemptiness problem}
of equations.  The nonemptiness problem of an
equation over a schema $\cal S$ with relation variables $\cal X$
is the problem of deciding, given a database over $\cal S$, whether
the equation has a solution on that database.
\emph{In the present section we will only consider equations that do not
contain equations inside.}

Let us begin by considering 
equations that are not necessarily sparse, but that still have only
flat variables.  The nonemptiness
problem of such a flat-variable equation is clearly in
NP\@.  Now suppose, moreover, that the database schema $\cal S$ is also flat;
then ${\cal S} \cup {\cal X}$ (the expansion of $\cal S$ with the relation
variables of the equation) is an entirely flat schema.
Of course, the equation $e_1 = e_2$ is still in general in the \emph{nested}
relational algebra, i.e., $e_1$ and $e_2$ can contain $\nu$ and $\mu$
operators.  A result by Paredaens and Van Gucht \cite{pvg_flatflat}, however,
implies that the nested relational algebra condition $e_1 = e_2$ can also be
expressed in the form $e \neq \emptyset$, with $e$ a \emph{flat} relational
algebra expression.  The nonemptiness problem of the equation
thus amounts to asking whether $\{(X_1,\dots,X_p) \mid e \neq
\emptyset\}$ is nonempty 
on a given database $B$ over $\cal S$.  Equivalently, we ask whether the
existential second-order logic ($\exists \rm SO$) sentence $\exists X_1 \dots
\exists X_p \, \varphi_e$ is true on $B$, where $\varphi_e$ is a first-order
logic sentence expressing that $e \neq \emptyset$.  Moreover, by the
equivalence of relational algebra and first-order logic, \emph{any}
$\exists \rm SO$ property can be obtained in this way.
Now, Fagin's theorem
\cite{fagin_theorem,ef_fmt} states that $\exists \rm SO$ captures
exactly the NP properties of flat relational databases.
Hence, \emph{the class of
nonemptiness problems of flat-variable
equations over flat database schemas is exactly
the class of NP properties of flat relational databases.}

What if $\cal S$ is not necessarily flat?
We next show that we still get exactly NP\@.
In essence, this is an extension of
Fagin's theorem to nested relational databases.

\begin{proposition} \label{propfagin}
Every property of nested relational databases
over some fixed schema $\cal S$, that is in NP and closed under isomorphism,
corresponds to the nonemptiness problem of some flat-variable
equation over $\cal S$.
\end{proposition}
\begin{proof}
The crux is a representation of nested relational databases by ``pseudo-flat''
ones, also used by Gyssens, Suciu, and Van Gucht \cite{gsv_restricted}.
Given a nested relational database $B$, we define its \emph{extended domain,}
denoted by ${\rm edom}(B)$,
as the union of its finite domain of atomic values with the set of all
relations occurring (possibly deeply nested) in $B$.  We regard the relations
in the extended domain as if they were atomic values.  Now for any nested
relational database schema $\cal S$ we can construct a flat one $\Bar
{\cal S}$, together with a mapping $\it rep$
from the set of databases over $\Bar {\cal S}$ \emph{onto} the set of
databases over $\cal S$,
expressible in the nested relational algebra.
The details of this mapping need not concern us
here.  Important is that we can furthermore
construct a converse mapping $\it flat$
from the set of databases over
$\cal S$ to the set of databases over $\Bar {\cal S}$,
also expressible in the nested relational algebra,
with the 
following properties for each database $B$ over $\cal S$:
(1)~the finite domain of ${\it flat}(B)$ equals ${\rm edom}(B)$; and (2)~${\it
rep}({\it flat}(B)) = B$.  Note that while
${\it flat}$ is expressed in the nested relational algebra, the result ${\it
flat}(B)$ is not really a flat database, because of the nested relations
in the extended domain.
However, it is ``pseudo-flat,'' in the sense that 
we regard these relations as if they were atomic values.
For any relation name $R$ of $\Bar {\cal S}$, we denote the 
nested relational algebra expression defining the $R$-component of the mapping
$\it flat$ by ${\it flat}_R$.  Likewise, we denote the expression defining the
$D$-component by ${\it flat}_D$.

Given this representation, the proof is straightforward.  Let $L$
be an NP property of databases over $\cal S$, closed under isomorphism.
Define the property $\bar L$ of databases over
$\Bar {\cal S}$ as follows: $F$ satisfies $\bar L$ if
${\it rep}(F)$ satisfies $L$.  Then $\bar L$ is in NP, and is
also closed under isomorphism.  Hence, Fagin's theorem gives us an
$\exists \rm SO$ sentence $\exists X_1 \dots \exists X_p \, \varphi$
over $\Bar {\cal S}$ expressing $\bar L$.  By the equivalence
of relational algebra and first-order logic, there is a flat relational
algebra expression $e$ over $\Bar{\cal S} \cup \{X_1,\dots,X_p\}$ such
that the first-order logic sentence
$\varphi$ is equivalent to $e \neq \emptyset$.  Now modify $e$ as
follows: for every relation name $R$ of $\Bar {\cal S}$, replace every
occurrence of $R$ in $e$ by ${\it flat}_R$.  Likewise, replace every
occurrence of $D$ in $e$ by ${\it flat}_D$. Denote the resulting nested
relational algebra expression by $e'$.

We now have, for any database $B$ over $\cal S$, that $B$ satisfies $L$
if and only if $\exists X_1 \dots X_p \, e' \neq \emptyset$ is true on $B$.
The condition $e' \neq \emptyset$ can easily be written as an equation
(cf.~Remark~\ref{disequation}).
\end{proof}

We are now ready to turn to sparse equations.  Their nonemptiness
problem is not
just in NP, but actually in
the complexity class
\emph{FewP} \cite{allender_sparsep},
consisting of all problems that can be decided
by a polynomial-time non-deterministic
Turing machine that has at most polynomially many accepting computations on
each input.  Clearly, ${\rm P} \subseteq {\rm FewP} \subseteq {\rm NP}$, but
the strictness of these inclusions remains open.

The obvious question to ask is whether Proposition~\ref{propfagin} remains
true if we focus on sparse equations, and replace `NP' by `FewP.' 
The answer is an easy ``yes,'' but then we must restrict attention to
\emph{ordered} databases: databases that include a total order on their
finite domain as one of their relations.
\begin{proposition} \label{propfaginfew}
Every property of ordered nested relational databases
over some fixed schema $\cal S$, that is in FewP and closed under isomorphism,
corresponds to the nonemptiness problem of some sparse
equation over $\cal S$,
when restricted to ordered databases only.
\end{proposition}
\begin{proof}
The usual proof of Fagin's
theorem immediately yields the case where $\cal S$ is flat.
Indeed, in that proof, to express an NP property decided by some
polynomial-time bounded
non-deterministic Turing machine $M$, one writes an $\exists{\rm SO}$ sentence
$\exists X_1\exists X_2\dots\exists X_p\,\varphi$ where
$X_1$ stands for
an order on the
domain; $X_2$, \dots, $X_p$ encode (using the order in $X_1$) a
computation of $M$; and $\varphi$ checks whether the computation is accepting.
As we are dealing with a FewP property,
$M$ has only polynomially many accepting computations.
Hence, the equation $\{(X_1,\dots,X_p) \mid \varphi\}$
would be sparse were it not for $X_1$, as there are exponentially
many possible orders on a finite domain.  On ordered databases, however, there
is no need for $X_1$ and we obtain a genuinely sparse equation.

This is for flat databases; for general nested
relational databases we use the same representation technique as in the proof
of Proposition~\ref{propfagin}.
\end{proof}

As a corollary we get

\begin{corol}
There exists a sparse equation whose nonemptiness problem is NP-complete, if
and only if\/ ${\rm FewP} = {\rm NP}$.
\end{corol}

\section{Sparse equations versus sparse powerset expressions}
\label{secresults}

Naturally, we call an equation algebra expression sparse if all equations
occurring in it are sparse. 
Inspired by Proposition~\ref{propsparse},
we can also define a sparsity condition on
powerset algebra expressions:
call a powerset algebra expression \emph{sparse} if its natural evaluation
strategy (defined in the obvious way) runs in polynomial space.

\begin{remark}
Using standard techniques one can show that sparsity is undecidable, for
equation algebra expressions as well as powerset algebra expressions.
An interesting question, raised by an anonymous
referee, is whether one can give useful
syntactic restrictions that guarantee sparsity.  Ideally every sparse
expression would be equivalent to one satisfying the syntactic restrictions.
\qed
\end{remark}

Suciu and Paredaens \cite{sp_powerset} showed that transitive closure
of a flat binary relation
is not expressible by a sparse powerset expression.  In
Example~\ref{exexpressions}, we gave an obvious
equation algebra expression for
transitive closure, but that expression was not sparse.  We can do better:

\begin{proposition} \label{proptransitive}
Transitive closure of a flat relation is expressible by a sparse equation
algebra expression.
\end{proposition}
\begin{proof}
Given a binary relation $R$ and a natural number $n \geq 1$, we define the 
relation $R^n$ as $R \circ \dots \circ R$ ($n$ times $R$),
where $\circ$ is the
classical composition operator of binary relations: $S \circ T = \pi_{1,4}
\sigma_{2=3} (S \times T)$.  Further, define $R^{\leq n}$ as $\bigcup_{i=1}^n
R^i$, and define $R^{=n}$ as $R^n - R^{\leq n-1}$.  Note that
$R^{\leq |R|}$ equals the transitive closure of $R$, and that 
for $n > |R|$, $R^{\leq n} = R^{\leq |R|}$.

Now consider the following 6-ary relation $\it Run$:
$$ {\it Run} :=
\bigcup_{i=1}^{|R|} R^{\leq i} \times R^{\leq i+1} \times R^{=i+1}. $$

We show next
that there is an equation whose \emph{only} solution, given $R$,
is $\it Run$.
This proves the Proposition, because all we then have to do is unnest
the solution set and project on the middle two columns to get the transitive
closure.  (The only exception is when $\it Run$ is empty, in which case the
transitive closure of $R$ is $R$ itself, but this can also easily be tested in
the nested relational algebra.)

The desired equation expresses the conjunction of the following conditions
on relation variable $X$:
\newcommand{\Xhh}{\check X}
\begin{enumerate}
\item
\label{uno}
For any pair $(x_5,x_6) \in \pi_{5,6}(X)$, we denote the relation
$$ \{(x_1,x_2,x_3,x_4) \mid (x_1,\dots,x_6) \in X\} $$
by $\tilde X(x_5,x_6)$, and denote further
\begin{align*}
\Hat X(x_5,x_6) & := \pi_{1,2}(\tilde X(x_5,x_6)) \quad \text{and} \\
\Xhh(x_5,x_6) & := \pi_{3,4}(\tilde X(x_5,x_6)).
\end{align*}

Then for every $(x,y) \in \pi_{5,6}(X)$, we must have
\begin{enumerate}
\item
\label{unoa}
$\tilde X(x,y) = \Hat X(x,y) \times \Xhh(x,y)$;
\item
\label{unob}
$\Hat X(x,y) \supseteq R$;
\item
\label{unoc}
$\Xhh(x,y) = \Hat X(x,y) \cup \Hat X(x,y) \circ R$; and
\item
\label{unod}
$(x,y) \in \Xhh(x,y) - \Hat X(x,y)$.
\item
\label{unoe}
Furthermore,
\emph{every} pair $(x',y')$ in the latter
difference belongs to $\pi_{5,6}(X)$,
with $\Hat X(x',y') = \Hat X(x,y)$ (and thus also $\Xhh(x',y') =
\Xhh(x,y)$).
\end{enumerate}
\item
\label{due}
$R^{=2} \subseteq \pi_{5,6}(X)$, and for every $(x,y) \in R^{=2}$, we have
$\Hat X(x,y) = R$.
\item
\label{tre}
For every $(x,y) \in \pi_{5,6}(X)$ such that $\Xhh(x,y) \circ R - \Xhh(x,y)
\neq \emptyset$, there exists a pair $(x',y') \in \pi_{5,6}(X)$ with
$\Hat X(x',y') = \Xhh(x,y)$.
\item
\label{quattro}
For every $(x,y) \in \pi_{5,6}(X)$ such that $\Hat X(x,y) \neq R$, there
exists a pair $(x',y') \in \pi_{5,6}(X)$ with
$\Xhh(x',y') = \Hat X(x,y)$.
\end{enumerate}

The conjunction of the above conditions expresses that $X$ equals $\it Run$.
Indeed, by (\ref{due}), (\ref{unoc}) and (\ref{unoa}) we know that
$R^{\leq 1} \times R^{\leq 2} \times R^{=2} \subseteq X$. By induction and by
(\ref{tre}), (\ref{unoc}), (\ref{unoe}) and (\ref{unoa}) we know that $R^{\leq
i} \times R^{\leq i+1} \times R^{=i+1} \subseteq X$ and hence ${\it Run}
\subseteq X$. Moreover, for every $(x,y) \in R^{=i+1}$ we have
$\{(x_1,y_1,x_2,y_2) \mid (x_1,y_1,x_2,y_2,x,y) \in X\} = R^{\leq i} \times
R^{\leq i+1}$. On the other hand, if $(x,y) \in \pi_{5,6}(X)$ then $\Hat
X(x,y) = R$, in which case $(x,y) \in R^{=2}$ by (\ref{unoc}) and
(\ref{unod}), or $\Hat X(x,y) \neq R$, in which case we know by induction and
by (\ref{quattro}) and (\ref{unob}) that $(x,y) \in R^{=i+1}$ for some $i$.
This proves $X = {\it Run}$.
\end{proof}

\begin{remark}
The equation constructed in the above proof is not only sparse, it is
\emph{unambiguous:} it has a unique solution on each input database.
Moreover, the same proof works more generally for any \emph{fixpoint
query} \cite{ahv_book} on flat databases. The only difficulty
is that fixpoint queries start from the empty relation, while in our
proof of Proposition~\ref{proptransitive} we start from $R$, but that
is easily dealt with.  As already explained in the Introduction, we
thus basically rediscovered an earlier result by Kolaitis to the effect
that every fixpoint query is implicitly definable in first-order logic
\cite{kolaitis_implicit}. But note the directness of our proof,
straightforwardly specifying the run of the fixpoint
computation in an unambiguous way.  The original proof
(also presented by Ebbinghaus and Flum \cite{ef_fmt}) is a bit more
roundabout, specifying the ``stage comparison'' relation instead.
\qed
\end{remark}

Another, perhaps a bit frivolous,
example of a query that is expressible using sparse
equations but not using sparse powerset expressions is the nesting operator
$\nu$.  It is easy to express $\nu$ in the powerset algebra using the powerset
operator and the other operators, but not $\nu$ itself; so $\nu$ is not
primitive in the powerset algebra.
As a consequence (Proposition~\ref{proppowerset}), $\nu$ is not
primitive in the equation algebra either.  We next observe that
when we restrict to sparse expressions, nesting remains imprimitive in the
equation algebra, but becomes primitive again in the powerset algebra.

\begin{proposition} \label{notinpowerset}
Nesting is not expressible by a sparse powerset expression without using the
$\nu$ operator itself.
\end{proposition}
\begin{proof}
Suppose we want to express nesting of a flat binary relation $R$.
The first application of the powerset operator is to the result of a flat
relational algebra expression $e$ applied to $R$.  Let us focus on
the case where $R$ is the identity
relation on a finite domain of $n$ elements.
A straightforward argument by structural induction
shows that, on identity relations, every relational algebra expression
is equivalent to a finite disjunction of \emph{equality
types}.  Here, an equality type is a maximally consistent conjunction of
equalities $x_i=x_j$ and non-equalities $x_i \neq x_j$ over the variables
$x_1,\dots,x_k$, where $k$ is the output arity of $e$.
We thus see that
either $e(R)$ is empty on all such $R$ (this is when the disjunction is empty),
or $e(R)$ is of size at least $n$ when $n$ is at least $k$.
In the empty case, the powerset operator is useless, and we
continue to the next application of powerset.
Otherwise, the powerset operator explodes and the
overall expression is not sparse.
\end{proof}

\begin{proposition}
Nesting is expressible by a sparse equation expression without using the $\nu$
operator itself.
\end{proposition}
\begin{proof}
Let $R$ be a relation name of type $(0,0)$, and let $X$ and $Y$ be
relation variables of type $(0)$.  We can write a relational algebra
expression $e$ such that on any database $C$ over $\{R,X,Y\}$,
$e(C)$ is empty if and only if $X$ is a singleton $\{x\}$ with $x \in 
\pi_1(R)$, and $Y = \{y \mid (x,y) \in R\}$.
An example of an $e$ that works is:
($\Delta$ stands for symmetric difference)
$$
\pi_1\sigma_{1\neq 2}(X \times X) \cup
(X - \pi_1(R)) \cup
(Y \mathbin{\Delta} \pi_3\sigma_{1=2}(X \times R))
$$
Hence, the expression
$$ \mu_1 \bigl ( \{(X,Y) \mid e = \emptyset\} \bigr ) $$
is a sparse equation expression equivalent to $\nu_2(R)$.

The construction for general nesting operations is analogous.
\end{proof}

Our final, and main technical, contribution concerns the parity query.
Suciu and
Paredaens showed that
the parity of the cardinality of a finite set is not expressible by a
sparse powerset expression.  We show the analogue for the equation algebra:

\begin{proposition}
The parity query is not expressible by a sparse equation expression.
\end{proposition}
\begin{proof}
Suppose, to the contrary, that we have a sparse equation expression to express
the parity of the cardinality of a finite domain $D$.  We may assume that
the input schema is empty, i.e., an input database consists of $D$ and nothing
else.  Consider an innermost equation $E_0$ occurring in our expression. It
may be nested inside other equations $E_1,\ldots,E_k$, enumerated from
the inside to the outside.
By Remark~\ref{disequation}, for each $j\in\{0,\ldots,k\}$, 
$E_j$ can be 
written in the form $\{(X^j_1,\ldots,X^j_{i_j}) \mid e_j \neq
\emptyset\}$, for some $i_j$, 
where $e_j$ is a flat relational algebra expression over the flat
schema $\{X^j_1,\ldots,X^j_{i_j}\}$ possibly expanded
with certain free variables $X^l_m$, where $l>j$ and $m\le i_l$.  

By our assumption there are at most polynomially many solutions to
equation $E_k$. For each solution $A_k$ of $E_k$ there are at most
polynomially many solutions $A_{k-1}$ of $e_{k-1}$ and so on. A sequence
$\bar A = (A_k,\ldots,A_0)$ of databases is a
{\em solution vector for $E_0$}, if
$A_k$ is a solution for $E_k$, and each $A_j$, $j<k$ is a solution for
$E_j$, given $A_k,\ldots,A_{j+1}$.  

Now let $\bar A= (A_k,\ldots,A_0)$
be a solution vector for $E_0$, given an input
$D$ of size $n$. Then for every permutation $f$ of $D$,
$f(\bar A)=(f(A_k),\ldots,f(A_0))$ is also a solution vector for
$E_0$.  The number of different such $f(\bar A)$ is precisely $n!/|{\rm
  Aut}(\bar A)|$, where ${\rm   Aut}(\bar A)$ is the group of
automorphisms of $\bar A$.  Since all equations
are supposed to be sparse, this number is at most $n^\ell$ for some
fixed $\ell$, or, 
equivalently, $|{\rm Aut}(\bar A)| \geq n!/n^\ell$.  Putting $k=\ell+1$, this
implies $|{\rm Aut}(\bar A)| \geq (n-k)!$ for sufficiently large $n$.  

We thus need to know more about large permutation groups.
The following crucial lemma
will give us the information we need.
The group of permutations of a finite set $D$ is denoted by ${\rm Sym}(D)$,
and its alternating subgroup of even permutations by ${\rm Alt}(D)$.
If $G$ is a subgroup of ${\rm Sym}(D)$, a \emph{fixed set} for $G$ is a subset
$\Delta \subseteq D$, such that every $g \in G$ maps $\Delta$ to $\Delta$.
The action of $G$ on a fixed set $\Delta$ (as a subgroup of
${\rm Sym}(\Delta)$) is denoted by $G^\Delta$.

\begin{lemma} \label{cruciallemma}
Let $k$ be a fixed natural number.
Let $G$ be a subgroup of ${\rm Sym}(D)$, $|D| = n$, $n$ sufficiently large.
Then $|G| \geq (n-k)!$ implies the existence of a fixed set $\Delta$
with $|\Delta| \geq n-k$, such that $G^\Delta$ contains ${\rm Alt}(\Delta)$.
\end{lemma}
\begin{proof}[Proof of Lemma~\ref{cruciallemma}]
For background on finite permutation groups, we refer to Wielandt's book
\cite{wielandt}, but here are a few preliminaries.
An \emph{orbit} of a permutation group $G$ on a set $D$ is
a set of the form $\{g(x) \mid g \in G\}$, for some $x \in D$.
We call $G$ \emph{transitive} if $D$ is one single orbit of $G$.
Further, $G$ is called \emph{primitive} if it is transitive and has no
nontrivial blocks.  Here, a \emph{block} of $G$ is a subset
$\Delta \subseteq D$ such that for all $g \in G$, the set $g(\Delta)$
is either equal to $\Delta$, or disjoint from it.  \emph{Trivial} blocks
are $\emptyset$, $D$, and the singletons.  If $G$ is not primitive but
transitive, there is always a \emph{complete block system} which partions
$D$ in equal-sized non-trivial blocks.  We recall:
\theoremstyle{plain}
\newtheorem*{BT}{Bochert's Theorem (1889)}
\begin{BT}
Let $G$ be primitive on $D$, not containing ${\rm Alt}(D)$.
Let $|D| = n$.  Then $|G| \leq n!/\lceil n/2 \rceil !$.
\end{BT}

For the proof of the Lemma, 
first assume that
$G$ is transitive.  There are two possibilities:
\begin{enumerate}
\item
$G$ is imprimitive with, say, $b$
blocks of size $a$ ($a>1$, $b>1$, $ab=n$).  Then $|G| \leq b! \, (a!)^b$,
which in turn is at most $2 ( \lfloor n/2 \rfloor ! )^2$ for $n$ sufficiently
large.  Thus, by what is given about $|G|$,
\begin{equation}
(n-k)! \leq 2 ( \lfloor n/2 \rfloor !)^2.
\label{imprimitive}
\end{equation}
However, this is impossible for $n$ sufficiently large.
\item
$G$ is primitive.  Then, unless $G$ contains ${\rm Alt}(D)$ (in which case the
Lemma is proved), by Bochert's theorem,
\begin{equation}
(n-k)! \leq \frac{n!}{\lceil n/2 \rceil !}.
\label{primitive}
\end{equation}
Again, this is impossible for $n$ sufficiently large.
\end{enumerate}

Now assume
$G$ is intransitive.  Let $\Delta$ be an orbit of $G$ of maximal size; let 
$\ell \geq 1$ be such that the size of $\Delta$
equals $n - \ell$.  We have $|G| \leq
(n - \ell)! \, \ell!$.
Suppose $\ell > k$.  Then $(n - \ell)! \, \ell!$
reaches its maximum at $\ell =
k+1$.  Hence, $(n-k)! \leq |G| \leq (n - k - 1)! \, (k+1)!$ and thus
$n-k \leq (k+1)!$ which is impossible for large enough $n$.

So, $\ell \leq k$, or in other words,
the size of $\Delta$ is at least $n-k$.
We have $|G| \leq \ell! \,
|G^\Delta| \leq k! \, |G^\Delta|$, so $|G^\Delta| \geq
(n-k)!/k!$.  By definition $G^\Delta$ is transitive on $\Delta$.
We can now apply the same arguments as in the case ``$G$ is transitive''
above, for $G^\Delta$ instead of $G$, and get that $G^\Delta$ must contain
${\rm Alt}(\Delta)$.  Indeed,
in the right-hand sides of inequalities \ref{imprimitive} and \ref{primitive},
$n$ now becomes $n - \ell$, which has for effect that
the upper bounds become smaller.  Hence, as
these inequalities already were impossible, they now become even more
impossible.  The extra factor of $1/k!$ in the left-hand sides does not
have a significant influence.
\end{proof}

Invoking this Lemma for $G = {\rm Aut}(\bar A)$, we get a fixed
set $\Delta$ of size at least $n-k$ such that any even permutation of $\Delta$
can be extended to an automorphism of $\bar A$.

Now let $X$ be one of the relation variables of an equation $E_j$,
of arity, say, $r$, and consider any $r$-tuple $t$
whose components are either the symbol $*$ or are in $D - \Delta$.
Let $r'$ be the number of components that are
the symbol $*$.
Further, let $\xi$ be an equality type of $r'$-tuples.  Denote by ${\it
Neighbors}^{\bar A}_X(t,\xi)$ the set of $r'$-tuples over $\Delta$ of
equality type 
$\xi$ such that, if we replace the $*$-components of $t$ by the components of
the $r'$-tuple (from left to right), we get a tuple in $X^{\bar A}$.
\begin{claim} \label{schwentick}
${\it Neighbors}^{\bar A}_X(t,\xi)$ is either empty, or consists of
\emph{all} $r'$-tuples over $\Delta$ of equality type $\xi$.
\end{claim}
\begin{proof}[Proof of Claim~\ref{schwentick}]
Suppose to the contrary that
$N := {\it Neighbors}^{\bar A}_X(t,\xi)$ is neither empty nor full.
Take $h_1$ in $N$, and take $h_2$ (of arity $r'$ and of type
$\xi$) not in $N$.
Take two arbitrary elements from $\Delta$ that neither appear in $h_1$ nor in
$h_2$, and remove them from $\Delta$, resulting in $\Delta'$.
Take a third tuple $h_3$ (of the right arity and
type) over $\Delta'$, and disjoint from $h_1$ and $h_2$.
If $h_3$ is in $N$, initialize the set
$I$ to $\{h_2,h_3\}$; otherwise, put $I := \{h_1,h_3\}$.
Now complete $I$ to
a maximal set of pairwise disjoint 
$r'$-tuples over $\Delta'$ of equality type $\xi$.
There are at least $(n-k-2)/r'$
tuples in $I$.

Assume at least half of $I$ is outside $N$;
denote the set of these by $I'$.  (The case where at least half
of $I$ is \emph{in} $N$ is symmetric.)
Fix an $h \in I \cap N$.
For each tuple $s$ in $I'$, we consider
the permutation $\varpi_s$ that transposes $s$ and $h$ and leaves
everything else fixed.  If $\varpi_s$ happens to be odd, we make it even
by adding the transposition of the two dummy elements we took out of
$\Delta$ (when we defined $\Delta'$).  Then each set $\varpi_s(N)$ contains
$s$, but does not contain any other tuples from $I'$.  Thus, we produce
in this way at least $f(n) := ((n-k-2)/2r'$ different sets of
$r'$-tuples over $\Delta$.  Since they are even, each $\varpi_s$
can be extended to an automorphism.  Hence, each of the $f(n)$ sets
must be the ${\it Neighbors}^{\bar A}_X(t',\xi)$ of some $t'$.  However, there
are less than $(k+1)^r$ different possibilities for $t'$, while $f(n)$
is larger than that for $n$ sufficiently large.  So we get to the
desired contradiction.
\end{proof}

We call $(t,\xi)$ an \emph{$r$-ary pattern.}
If ${\it Neighbors}^{\bar A}_X(t,\xi)$ is nonempty (and thus
full), we say that the pattern is \emph{instantiated in $X^{\bar A}$.}
Note also that the extreme cases, where $t$ consists exclusively of stars
or where $t$ has no star at all, are also allowed and make sense.

By the above, we thus see
that any solution vector
$\bar A$ can be generated by the following non-deterministic
procedure:
\begin{enumerate}
\item
Initialize all relations of $\bar A$ to empty.
\item
Choose at most $k$ different elements from $D$, playing
the role of the elements outside $\Delta$.
\item
For every relation variable $X$ (of arity $r$, say), run through all $r$-ary
patterns, and for each of them, non-deterministically instantiate it
in $X^{\bar A}$, or not.
\end{enumerate}
Since $k$ and the number of relation variables are
fixed, the number of possible patterns is also fixed.
Hence, we can write an expression in the nested relational
algebra that, given $D$, constructs the set of all possible 
vectors $\bar A$ of the above non-deterministic procedure.  This set
is a superset of the actual set of solution vectors for $E_0$. 
Equation $E_0$ can now be replaced by a nested relational algebra
expression which (1) constructs the set of solution candidates $\bar
A$, (2) projects out the relations for $X^0_1,\ldots,X^0_{i_0}$ and
(3) selects those relations which fulfil $E_0$. The latter is 
an easy task for the nested relational algebra \cite{gvg_powerset}.

Hence, we can get rid of $E_0$.  Repeating this
process, we can get rid of all equations, so that in the end we are
left with a standard nested relational algebra expression for the parity
query.  But this is well known to be impossible \cite{ahv_book,pvg_flatflat}.
\end{proof}

\begin{remark}
Note that we have actually shown that over the empty schema, where databases
consist of a finite domain and nothing else, the
sparse equation algebra is no more powerful than the standard
relational algebra.
As a matter of fact, the proof can easily
be generalized to apply also to schemas having only relation names of type
$(0)$.
\qed
\end{remark}

\section*{Acknowledgment}  
We are indebted to L\'aszl\'o Babai, who pointed us to Liebeck's paper.

\end{document}